\documentclass[12pt,preprint]{aastex}
\usepackage{psfig,lscape}
\def\lap{\lower.5ex\hbox{$\; \buildrel < \over \sim \;$}}
\def\gap{\lower.5ex\hbox{$\; \buildrel > \over \sim \;$}}

\def\ergcm2s{${\rm erg\ cm^{-2}\ s^{-1}}$}

\def\ergscm2s{${\rm erg\ cm^{-2}\  s^{-1}}$}

\def\cm-2{${\rm cm^{-2}}$}

\begin{document}

\title{Variable Star Candidates in an ACS Field of M31}

\author{Benjamin F. Williams\altaffilmark{1}}

\altaffiltext{1}{Harvard-Smithsonian Center for Astrophysics, 60
Garden Street, Cambridge, MA 02138; williams@head-cfa.harvard.edu}

\begin{abstract}

A search for variable stars is performed using two epochs of Hubble
Space Telescope (HST) Advanced Camera for Surveys (ACS) imaging data
for a 9.28 square arcminute portion of M31.  This data set reveals 254
sources that vary by at least 4$\sigma$ between epochs.  The positions
and 2-epoch $B$-band (equivalent) photometry of these sources are
presented.  The photometry suggests that this catalog includes most of
the RR Lyrae population of this portion of M31.

\end{abstract}

\clearpage

\section{Introduction}

The proximity of M31 has made it one of the most important galaxies
for the study of variable stars.  Cepheid studies date back more than
40 years \citep{baade1963,baade1965}, and continue in recent years.
For example, the DIRECT project has found hundreds of Cepheids, as
well as eclipsing binaries in M31
\citep{kaluzny1998,kaluzny1999,stanek1998,stanek1999,mochejska1999,bonanos2003,joshi2003}
to exploit the fact that M31 provides a crucial step in the
extragalactic distance ladder.

Variability studies of M31 have become diverse over last the few
decades.  These efforts include surveys for classical novae
(\citealp{rosino1973,rosino1989,sharov1991,shafter2001,filipenko2002,rector2003,darnley2004},
and many others) and most recently, efforts to detect microlensing
events
\citep{crotts1992,crotts1996,ansari1997,auriere2001,kerins2001,paulin2003,calchi2002,calchi2003,kerins2003,dejong2004}.

The data from these extensive surveys for Cepheids, novae, and
microlensing events has provided large catalogs of the bright variable
stars in M31.  The POINT-AGAPE collaboration has published a catalog
with 35,414 variable stars to an $r$ magnitude limit $\sim$23 in 0.6
square degrees, not overlapping with the field discussed herein. This
wide-field study determined that the variable star distribution is
highly asymmetric \citep{an2004}.  The DIRECT project (see references
above) cataloged hundreds of variable stars in 5 fields, each 120
square arcmin.  My ACS field overlaps with their field M31D
\citep{kaluzny1999}, but the ACS data is several magnitudes deeper.
\citet{joshi2003} cataloged hundreds more variable stars in a 170
square arcmin field, and \citet{mould2004} has completed a large
survey for bright, long period variables in M31, cataloging nearly
2000. Their fields also overlap with my ACS field, but are not as deep
as my ACS data.

The deepest ground-based searches for M31 variables to date have been
those of \citet{dolphin2004} and \citet{pritchet1987}.  These surveys
were sufficiently deep to study the RR~Lyrae
population. \citet{dolphin2004} cataloged 37 variable stars in a
search for RR~Lyrae stars in the M31 halo.  They found 24 RR~Lyrae
stars, whose magnitudes suggest that they are metal poor.  Their
analysis suggested an RR~Lyrae density of $\sim$1 arcmin$^{-2}$.
These results contrasted those of \citet{pritchet1987}, who found 30
RR~Lyrae stars in a 7 square arcminute field.  Their completeness
analysis implied $\sim$17 RR~Lyrae stars arcmin$^{-2}$.  The deepest
search to date is that of \citet{brown2004}, who have used $HST$/ACS
to search for variables in M31, finding 55 RR~Lyrae stars, implying a
density of $\geq$4 arcmin$^{-2}$ in their field.

Herein, I report the results of a search for variables in M31
with ACS imaging. Section 2 describes the 2-epoch ACS data set used
and the analysis performed.  Section 3 gives the results of the
analysis.  Section 4 discusses the types of variable stars these are
most likely to be, and \S 5 summarizes the conclusions.

\section{Data}

\begin{figure}
\centerline{\psfig{figure=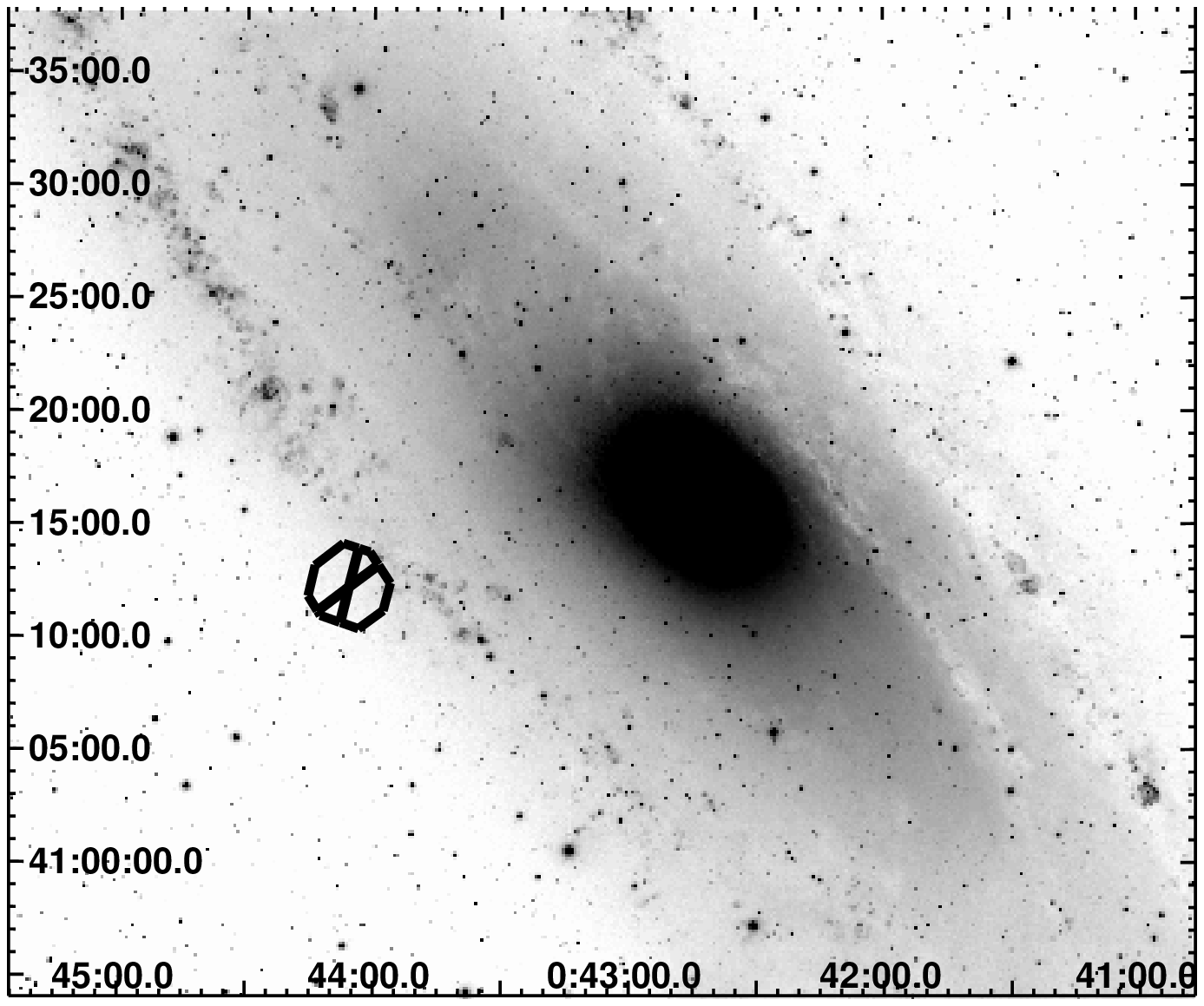,width=6in,angle=0}}
\caption{An H$\alpha$ image of a portion of M31 is shown.  The polygon
drawn in black at 00:44.1, 41:12.1 outlines the area where the ACS
data set was sensitive to variable sources.  The cross inside the
polygon shows the location of the chip gap in each of the 2 epochs.}
\label{survey}
\end{figure}

The data used for this project were originally obtained as part of a
program to find the optical counterparts for X-ray novae in M31
\citep{williams2005}.  I obtained two sets of {\it HST} ACS data, one
observed at UT 21:35 on 03-Dec-2003 and one observed at UT 16:38 on
01-Mar-2004.  Each of these were pointed at R.A.=00:44:07,
Dec.=41:12:19.5.  The observations had orientations of 71.75 deg and
32.73 deg respectively.  Both observations were taken using the
standard ACS box 4-point dither pattern to allow the final data to be
drizzled to recover the highest possible spatial resolution.  All
exposures were taken through the F435W filter.  The total exposure
times were 2200 seconds for each data set.  The 9.275 arcmin$^2$
overlapping region of the two ACS images is outlined on a wide-field
H$\alpha$ image of M31 in Figure~\ref{survey}.

\subsection{Image Processing, Alignment and Subtraction}

I aligned and drizzled each set of 4 images into high-resolution
(0.025$''$ pixel$^{-1}$) images using the PyRAF\footnote{PyRAF is a
  product of the Space Telescope Science Institute, which is operated
  by AURA for NASA.} task {\it multidrizzle},\footnote{multidrizzle is
  a product of the Space Telescope Science Institute, which is
  operated by AURA for
  NASA. http://stsdas.stsci.edu/pydrizzle/multidrizzle} which has been
optimized to process ACS imaging data.  The task removes the cosmic
ray events and geometric distortions, and it drizzles the dithered
frames together into one final photometric image with pixel values in
units of counts per second.

I aligned the final photometric ACS images with the coordinate system
of the Local Group Survey (LGS; \citealp{massey2001}) by applying the
IRAF\footnote{IRAF is distributed by the National Optical Astronomy
Observatory, which is operated by the Association of Universities for
Research in Astronomy, Inc., under cooperative agreement with the
National Science Foundation.} task {\it ccmap} to the centroids of
stars and globular clusters common to the ACS and LGS images.  The LGS
images have an assigned J2000 (FK5) world coordinate system accurate
to $\sim$0.25$''$, and they provided the standard coordinate system to
which I aligned all of the data for this project.  The alignment
errors measured by {\it ccmap} were $\sim$0.025$''$, or one pixel in
the drizzled images.

In order to produce ACS images from each epoch with aligned pixel
orientations, I used the IRAF tasks {\it geomap} and {\it geotran} to
map and transform the images so that the pixels were aligned in the
North up, East left orientation.  I used the same objects to
transform both epochs to the new pixel orientation.  This technique
resulted in images with identical pixel orientations and aligned to an
accuracy of 0.2 pixels.

Since the aligned data from both epochs had the same exposure times,
were taken through the same filter, and were processed with identical
techniques, the images were subtracted directly to search for residual
sources that were due to their brightness changing between the two
observations.

\subsection{Search Technique}

Most bright stars left some form of residual in the difference image
due to the different roll angles of the two fields and different
locations of the stars on the ACS detector.  Although these residuals
had a distinctive appearance that did not resemble a point source,
they severely hampered automated searches of the difference image by
producing an overwhelming number of false variability detections.  As
these images are very large (10$^{8}$ pixels), automated search
routines proved slow and inefficient.  Preliminary visual searches of
the difference image greatly reduced the number of false variability
detections, increasing the efficiency and reliability of the search,
with 93\% of the visual candidates passing the remaining tests for
variability between the two epochs (see \S~\ref{phot}).

\begin{figure}
\centerline{\psfig{figure=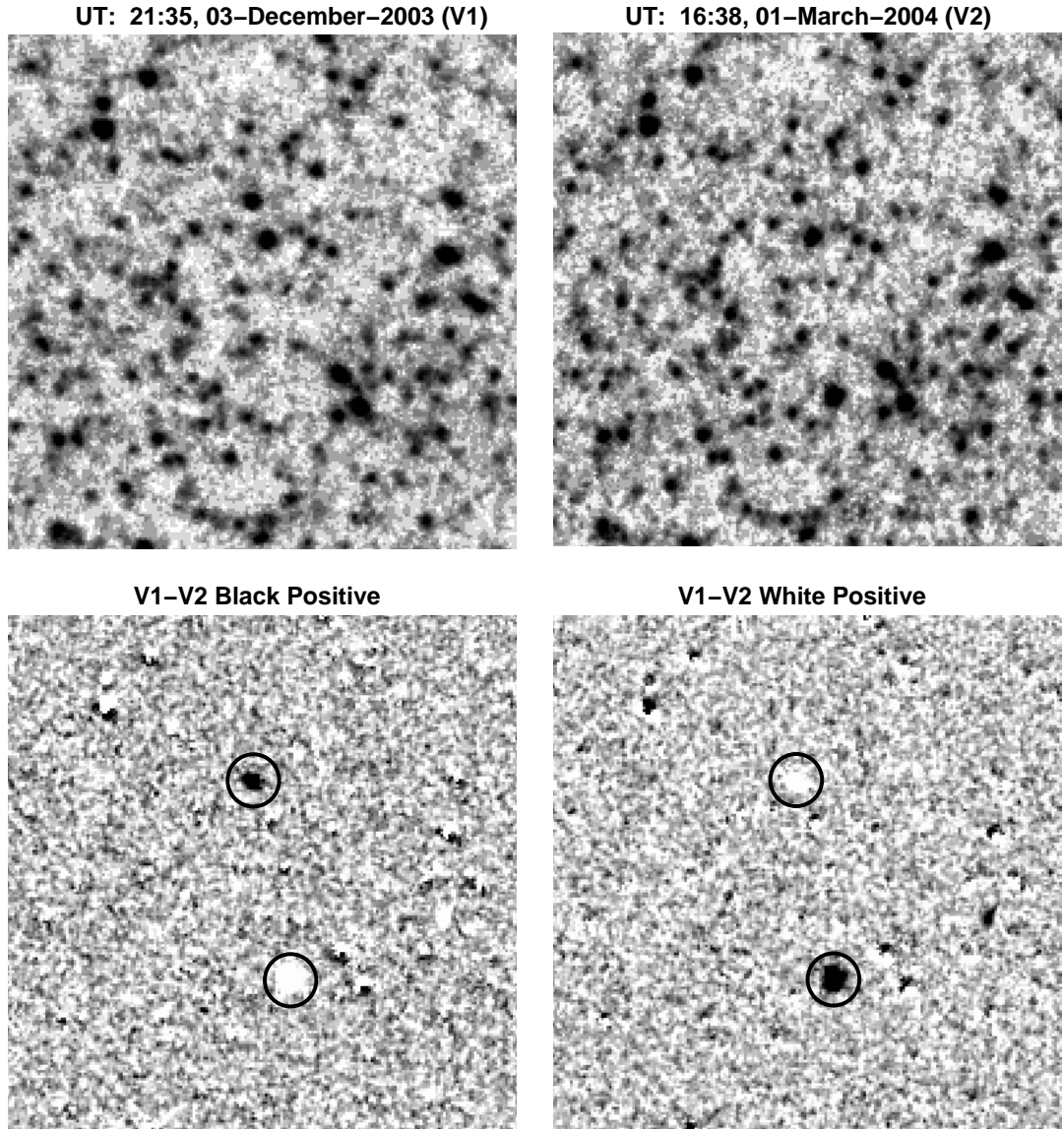,height=6in,angle=0}}
\caption{Example images from the variable search.  {\it Upper left:} A
$5'' \times 5''$ portion of the first ACS image of the field, taken
03-December-2003, designated as V1; black is positive. {\it Upper
right:} A $5'' \times 5''$ portion of the second ACS image of the
field, taken 01-March-2004, designated as V2; black is positive. {\it
Lower left:} The difference image, V1$-$V2, displayed so that black is
positive, revealing variables like M31ACSV J004407.7+411333 (marked
with the upper black circle), that were brighter during the first
exposure. {\it Lower right:} The difference image, V1$-$V2, displayed
so that black is negative, revealing variables like M31ACSV
J004407.7+411331 (marked with the lower black circle), that were
brighter during the second exposure.}
\label{search}
\end{figure}

The difference image was searched twice visually for variable
candidates.  All residuals that appeared as point spread functions
were flagged.  The first search was performed with the contrast of the
image set so that sources that were brighter during the first epoch
appeared as black spots.  The second search was performed with the
inverse contrast setting, in order to find objects that were brighter
in the second epoch.  Examples of a variable that became brighter and
one that became fainter, including their appearances at both contrast
settings, are shown in Figure~\ref{search}.

The searches resulted in 141 candidates that were brighter in the
first epoch and 133 candidates that were brighter in the second epoch.
The fact that these numbers were within $\sqrt{N}$ of one another was
encouraging that the search technique was not introducing any large
bias to the sample.  

The candidates' final positions were determined with the IRAF task
{\it imcentroid}.  Positions of the objects were measured with data
from the epoch in which the object appeared brighter.  Two objects did
not provide reliable centroids, suggesting that they were not strong
detections and/or not point sources.  These candidates were removed,
leaving a final list of 272 variable star candidates to be measured.

\subsection{Photometry}\label{phot}

Aperture photometry was performed at the {\it imcentroid} position for
each candidate using the IRAF task {\it phot}.  The aperture radius
was 0.15$''$, and the annulus used to measure the sky brightness was
from 0.30$''$ to 0.55$''$.  The photometry was performed identically
for both epochs.  These measurements yielded the ACS count rate for
each candidate during each observation epoch.

The count rates measured from the ACS images were converted to VEGA
magnitudes using the calibration provided in the ACS Data
Handbook\footnote{http://www.stsci.edu/hst/acs/documents/handbooks/DataHandbookv2/ACS\_longdhbcover.html}.
The VEGA magnitudes for the F435W filter is the ACS equivalent to the
Johnson $B$ magnitude.  The final conversion formula was:
$$
m_B \approx F435W_{VEGA} = -2.5 log(r) + 25.76 
$$ 
where $r$ is the count rate measured by {\it phot}.  Sources with
$B\sim27.8$ were detected with 5$\sigma$ significance.  I took this to
be the detection limit, only accepting detections brighter than this
and assigning an upper-limit $B>27.8$, to all non-detections.

The final $B$ magnitudes from the second epoch were subtracted from
those of the first, and the photometric errors were added in
quadrature.  No errors were assigned to non-detections.  

All of the candidates whose magnitudes changed by four times the
root-sum-square of their errors (or more) were included in the final
catalog of variables given in Table~\ref{cat}.  This requirement
eliminated 3 candidates from the original list, reducing the catalog
to 269 variable stars.  Finally, for all cases of stars detected in
only one epoch, the four dithered ACS exposures of the detection epoch
were inspected to ensure that these detections were not cosmic-rays or
artifacts from the image processing. Real stars were seen in all of
the individual raw exposures, while cosmic rays and artifacts were
not. These inspections found 15 of the variable candidates to be
spurious: cases where {\it multidrizzle} had failed to properly reject
a cosmic ray or artifact.  The final catalog therefore contains 254
candidate variable stars.

The photometric properties of these variables are shown in
Table~\ref{phottab} and Figure~\ref{bdb}.  They follow the expected
pattern, as the smallest detected change in magnitude increases with
increasing $B$ magnitude.  These statistics provide further
confirmation that the search technique did not introduce any
significant bias into the sample.

Comparisons of this catalog with the three previous overlapping
surveys show that five variables from the \citet{kaluzny1999} catalog
and 6 variables from the \citet{joshi2003} catalog lie within the 9.28
arcmin$^2$ variable-sensitive portion of my data set.  Two of the six
\citet{joshi2003} survey variables were identified here as well, and
all five of the \citet{kaluzny1999} variables were identified here.
Four of the \citet{joshi2003} variables were not seen (V346, V351,
V352 and V354), and five of the \citet{mould2004} variables were not
seen (44001, 44008, 44020, 44044, and 44045), possibly because of the
differing bandpasses of their surveys which were completed in $R$ and
$I$.  On the other hand, all of the variables of \citet{kaluzny1999}
were independently identified here.  One source (DIRECT V4599 M31D)
was culled from my catalog when {\it imcentroid} did not converge on
its center in the ACS image, suggesting that it is blended or
extended. The six previously-cataloged variables, including names and
photometry from the previous identifications, are provided in
Table~\ref{previous}.

\section{Results}

\begin{figure}
\centerline{\psfig{figure=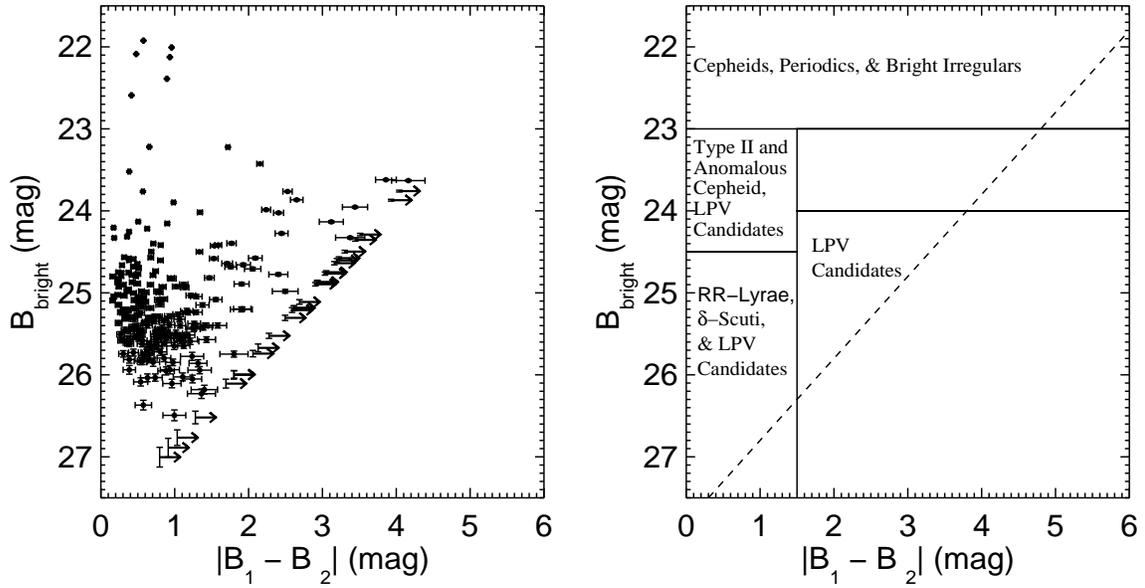,width=6.5in,angle=0}}
\caption{{\it Left:} The change in $B$ magnitude observed is plotted
against the brightest $B$ magnitude observed.  There is a dense
grouping of sources with characteristics typical of RR~Lyrae stars
($|\Delta B| \lap 1.5$ and $B \sim 25.5$). {\it Right:} The $|\Delta
B| - B$ plane is shown divided into labeled areas (solid lines); each
label gives the type of variable star that likely occupies that area
of the $|\Delta B| - B$ plane.  Eclipsing binaries could lie anywhere
in the $|\Delta B| - B$ plane.  The dashed line marks the magnitude
limit of this survey.}
\label{bdb}
\end{figure}

\begin{figure}
\centerline{\psfig{figure=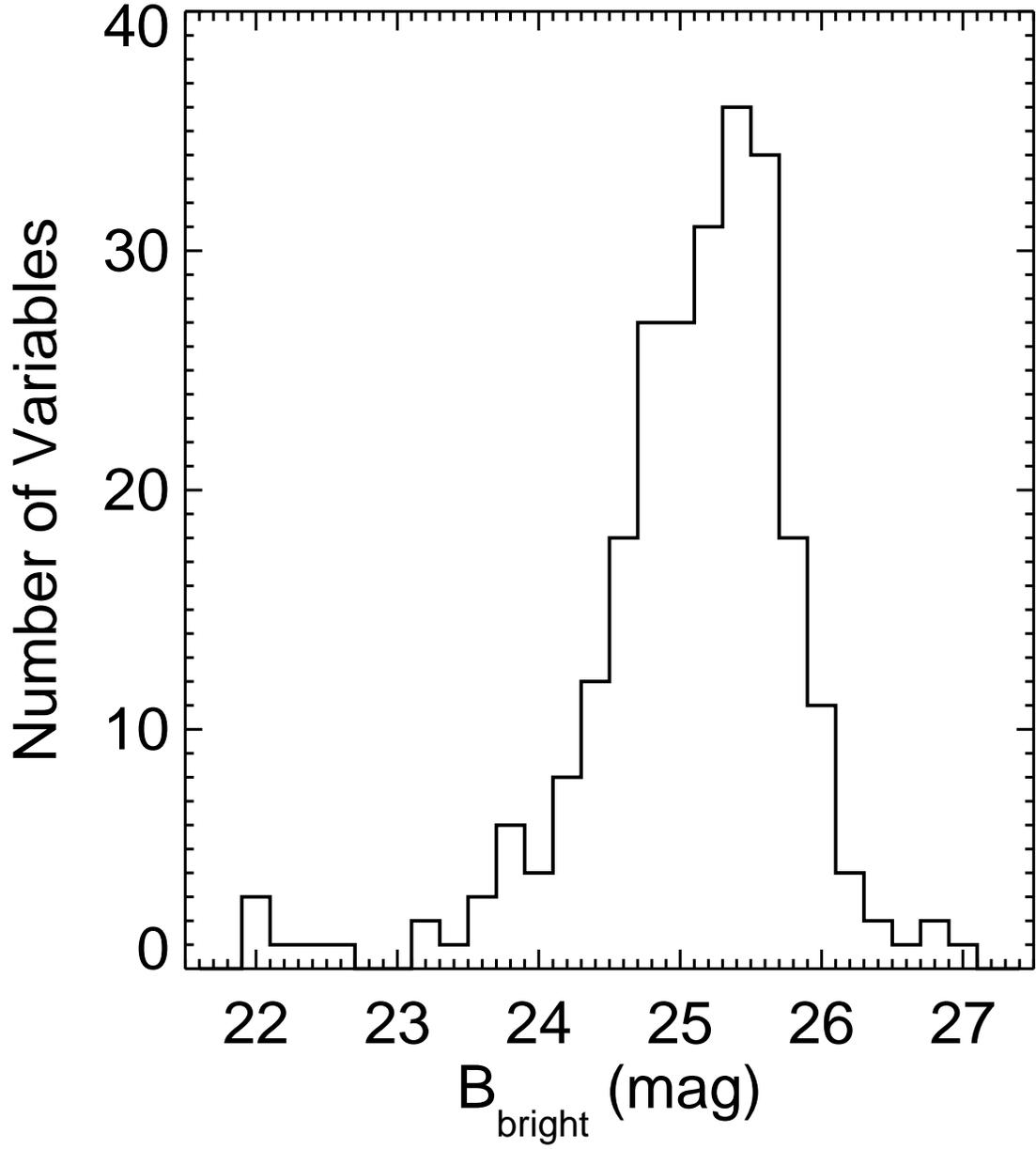,height=7in,angle=0}}
\caption{Histogram of the brightest $B$ magnitude observed vs. the
number of variables found.  The search technique fails for variables
that were not brighter than $B\sim26$ in this data set.}
\label{bn}
\end{figure}

\begin{figure}
\centerline{\psfig{figure=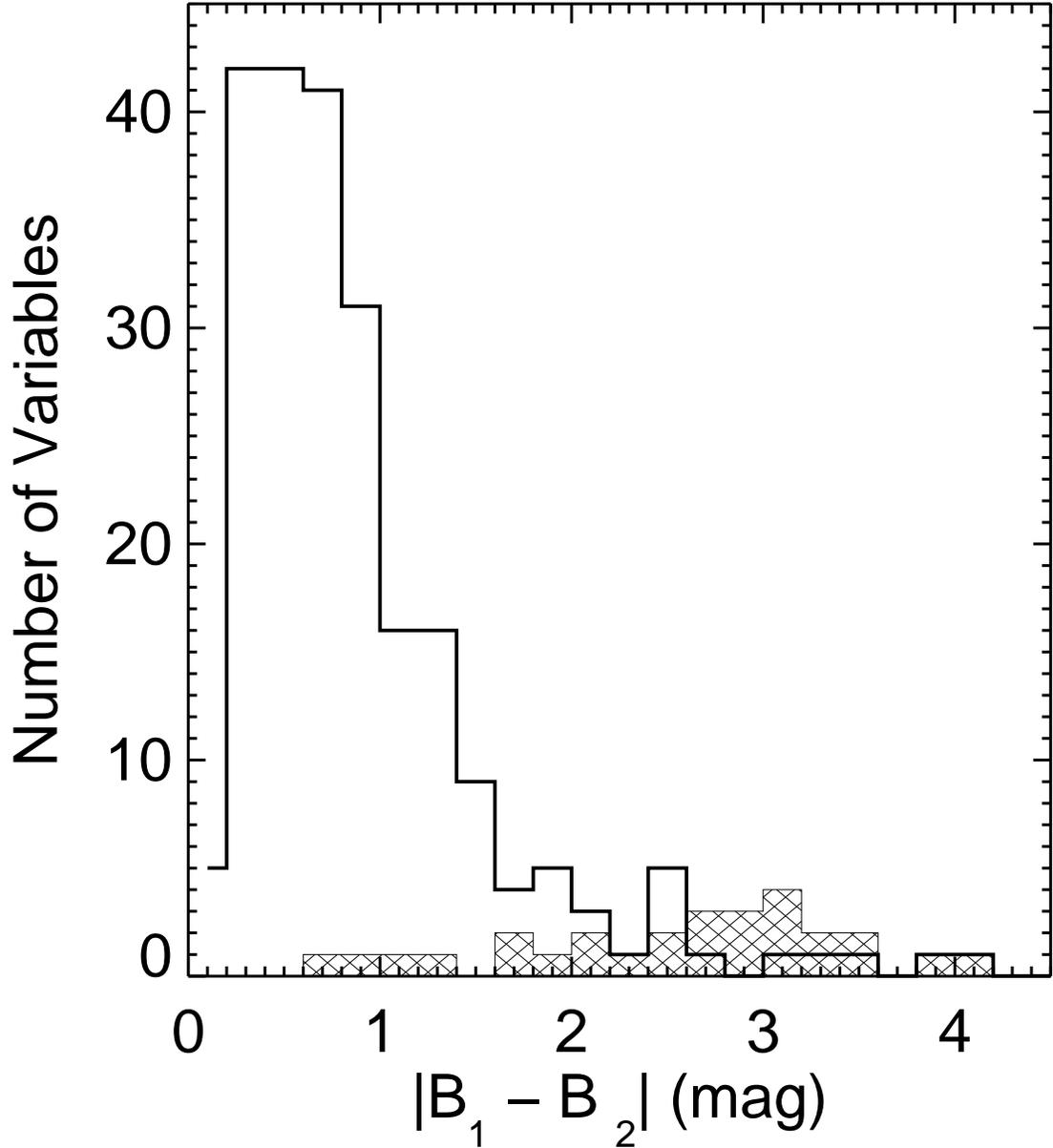,height=7in,angle=0}}
\caption{{\it Thick histogram:} the change in $B$ magnitude observed
vs. the number of variables found for the 226 variables detected in
both epochs. A clear peak is seen at $|\Delta B|\sim0.5$.  {\it
Hatched histogram:} the minimum change in $B$ magnitude observed
vs. the number of variables found for the 28 variables only detected
in one epoch.}
\label{dbn}
\end{figure}

The final catalog of 254 variable stars in these 9.28 arcmin$^2$ of
M31 is shown in Table~\ref{cat}.  Names were determined using the an
IAU-approved acronym (M31ACSV) followed by truncated J2000 coordinates
in an IAU-approved format (JHHMMSS.s+DDMMSS).  The variables range in
brightness from $21.9 < B < 27.0$, and they show brightness changes
$|\Delta B|>0.15$ mag.  The distribution of brightnesses and
brightness changes are shown in Figures~\ref{bdb}-\ref{dbn} and
discussed in \S~\ref{discussion}.  One important note about
Figure~\ref{bdb} is that, even though all of the brightness changes
seen in this survey are lower limits, only those that were not
detected in one of the two observations are marked as lower limits.

The density of variable stars in this small region of M31 is at least
27 arcmin$^{-2}$.  This is about twice as high as that seen by
\citet{brown2004} in a field 51$'$ from the nucleus.  The higher
density is not unexpected, as the field observed is only 17$'$ from
the nucleus.

This sample likely contains many types of variables, but only those
that vary by more than 0.15 mag.  Some could be novae, but as the
observed nova rate for the entire Galaxy is only $\sim25$ yr$^{-1}$
\citep{ramano1999}, clearly the bulk of the variables in this tiny
fraction of M31 are not novae.  Most of these variables are likely
pulsating stars and eclipsing binaries.  Pulsating variables in this
magnitude range that exhibit high-amplitude variability include
Cepheids, irregular variables, Type II and Anomalous Cepheids,
RR~Lyrae stars, high-amplitude $\delta$ Scuti stars (HADS), and Long
Period Variables (LPVs; \citealp{gautschy1996} and references
therein).  Well-sampled lightcurves will be needed to obtain reliable
source classifications; however, assuming eclipsing binaries randomly
populate the $|\Delta B| - B$ plane in low numbers, the relative
numbers of some types of variable stars can be roughly estimated based
on the $|\Delta B| - B$ diagram.

\section{Discussion}\label{discussion}

The right panel of Figure~\ref{bdb} shows the $|\Delta B| - B$ plane
divided into areas, each labeled with the type of variable star that
most likely occupies it.  Eclipsing binaries can lie anywhere in the
$|\Delta B| - B$ plane.  While these labeled areas are not reliable
classifications, they provide a guide regarding where certain types of
stars most likely fall in the figure.  This section discusses the
justification of these labels.

\subsection{The Brightest Candidates}

The six stars that reached $B < 23$ (J004400.3+411309,
J004401.0+411132, J004402.5+411137, J004404.0+411353, J004408.0+411243
and J004408.4+411347) would be easily detectable in ground based
images.  Four of these stars (J004401.0+411132, J004404.0+411353,
J004408.0+411243 and J004408.4+411347) were previously cataloged (see
Table~\ref{previous}).  The other two (J004400.3+411309 and
J004402.5+411137) changed in brightness by less than 0.5 mag, which
could be difficult to detect in ground-based photometry.  Fainter
variables would be quite difficult to detect in ground-based surveys
because of the effects of crowding.  The two other variables that were
previously cataloged were observed in the $R$ and $I$ bands, where
they have $m<21$ \citep{joshi2003}.

Those stars with $B_{bright} < 24.5$ are brighter than RR~Lyrae stars
in M31.  These stars may be Type II Cepheids and/or Anomalous Cepheids
\citep{wallerstein1984,wallerstein2002,marconi2004}.  These types of
Cepheids are typically fainter than Type I Cepheids and brighter than
RR~Lyrae stars.  Their amplitudes of variation are similar to RR~Lyrae
stars.  The catalog contains 16 newly-discovered variable candidates
that may have these characteristics (23$<B_{bright}<$24.5; $|\Delta B|
< 1.5$).  Although some of these stars could be bright LPVs, the
magnitudes of these stars in this study make them the best Type II and
Anomalous Cepheid candidates in the catalog.

\subsection{Fainter Candidates}

The histogram in Figure~\ref{bn} exhibits a clear peak at $B \approx
25.5$.  The density of variables appears to increase sharply at
magnitudes fainter than $B \approx 24.5$ or M$_B \approx -0.4$.  The
density increases from 2.2 arcmin$^{-2}$ mag$^{-1}$ to 14.6 arcmin$^{-2}$
mag$^{-1}$ as $B$ goes from 24 to 25. This increase is in proportion
to the increase in the stellar density, which goes from 220
arcmin$^{-2}$ mag$^{-1}$ to 1210 arcmin$^{-2}$ mag$^{-1}$ over the
same magnitude range.  These numbers show that at least $\sim$1\% of
the stars in this region of M31 with $B<25.5$ are variable.

Some of the decrease in the number of variables found at magnitudes
fainter than $B\approx 26$ is due to the limitations of the search
technique at those magnitudes on this data set.  Variables fainter
than $B\approx 26$ are more likely to be mistaken for noise in the
difference image, and as can be seen in Figure~\ref{bdb}, stars with
$B_{bright} > 26$ must change in brightness by $>$0.5 mag to pass all
of the variability criteria.

Although the peak in the brightness distribution can be explained by
completeness effects, it is interesting that the peak is at $B \approx
25.5$.  With a distance modulus of 24.47 and a typical extinction of
$A_B = 0.4$ mag \citep{williams2003}, this brightness corresponds to
M$_B \approx 0.6$, typical for RR~Lyrae stars with a range of
metallicities according to the luminosity--metallicity relation
\citep{chaboyer1999,cacciari2003}.  Allowing for the possibility that
RR~Lyrae stars in this field may cover a range in metallicities from
$-2\lap$[Fe/H]$\lap$0 (including disk and halo populations), they may
cover a range in luminosities from 0.3$< \langle M_V \rangle <$1.0.
The errors in the slope and zero point of the luminosity--metallicity
relation were included in the determination of this range. In
addition, RR~Lyrae stars could have amplitudes as great as $\Delta B
\approx 1.5$ (e.g., AQ Lyr; \citealp{castellani1998}).

The space density of RR~Lyrae stars in the solar vicinity is 6
kpc$^{-3}$ \citep{amrose2001}.  With an inclination of 77 degrees,
assuming a disk thickness of 1 kpc, 1 arcmin$^2$ of M31 corresponds to
$\sim$0.2 kpc$^3$.  Therefore, 1--2 RR~Lyrae arcmin$^{-2}$ from the
M31 disk should be seen.  If the RR~Lyrae density of the
\cite{brown2004} field ($\sim$4 arcmin$^{-2}$) is added to this disk
estimate, RR~Lyrae stars should account for {\it more than} 50--60 of
the variables in this catalog, as the present field is in a more dense
region of M31 than the \cite{brown2004} field.  Assuming foreground
extinction $A_B \approx 0.4$ and a range of $B-V$ colors from 0 to
0.5, RR~Lyrae stars are likely to occupy the $|\Delta B| < 1.5$ and
24.5$\lap B_{bright} \lap$26.5 portion of Figure~\ref{bdb}.  This
section of the plot contains 175 stars, suggesting that the catalog
includes a large fraction of the RR~Lyrae stars in this region of M31.

In addition, this section of the plot may contain some faint Type II
and Anomalous Cepheid stars, as well as bright HADS.  HADS are
distinct from RR~Lyrae stars in that they are typically Population I
and have shorter periods (see \citealp{gautschy1996,rodriguez2000} and
references therein).  These stars range in spectral type from A2--F0
(1.3$<$M$_B<$3.0), and can be as much as 2 magnitudes brighter than
the main sequence \citep{garcia1995}.  Therefore bright HADS can
overlap with the magnitude range covered by RR~Lyrae stars, adding to
the number of potential RR~Lyrae stars in the catalog.

Finally, many of these variable sources could be LPVs or eclipsing
binaries.  LPVs likely account for a large number of the variables in
the catalog, as \cite{brown2004} found them to be common in their ACS
field.  These stars have 1$\lap B-V\lap $3, can exhibit $B$ magnitude
variations of greater than 4 magnitudes, and have maximum luminosities
of $-1\lap M_B\lap 2$ \citep{celis1986,delaverny1997}. With only 2
epochs of photometry, LPVs are likely present at all B$_{bright}$
magnitudes fainter than $\sim$24, and the highest amplitude variables,
with $|\Delta B| > 1.5$ mag, are the most likely to be LPVs or
eclipsing binaries.

\section{Conclusions}

The 2-epoch $F435W$-band ($B$ equivalent) photometric properties of an
$HST$ ACS field in M31 have been measured, resulting in a catalog of
the positions and magnitudes of 254 likely variable stars in 9.28
arcmin$^2$.  This catalog provides a starting point from which to
study the variable star population of this region of M31 at RR~Lyrae
magnitudes.

With only 2 epochs of photometry, the variable sources cannot be
individually classified; however, with just two epochs there appears
to be a high density ($>$27 arcmin$^{-2}$) of variable stars in M31.
Four of the brightest variables are previously known periodic
variables, including one Cepheid.  The other bright variables could be
bright irregular, periodic, eclipsing binary, Type II Cepheid, or
anomalous Cepheid stars.

A comparison of the density of variables in this catalog at RR~Lyrae
magnitudes with the density of such stars in the Galaxy and in
previous M31 surveys suggests that this catalog includes most of the
RR~Lyrae stars in this portion of M31.  Additionally, the catalog
likely includes many LPVs, along with some high-amplitude $\delta$
Scuti stars and eclipsing binaries.

I thank Michael Garcia for his many helpful comments.  Support for
this work was provided by NASA through grant G0-9087 from the Space
Telescope Science Institute and through grant GO-3103X from the {\it
Chandra} X-Ray Center.


\clearpage

\clearpage

\clearpage

\clearpage

\clearpage

\clearpage
\oddsidemargin=-1cm
\tabletypesize{\scriptsize}

\begin{deluxetable}{cccccc}
\tablewidth{0in}
\tablecaption{Variable Stars in the ACS Field}
\tableheadfrac{0.01}
\tablehead{
\colhead{{Name (M31ACSV)}} &
\colhead{{R.A. (J2000)}} &
\colhead{{Dec. (J2000)}} &
\colhead{{$B_1$\tablenotemark{a}}} &
\colhead{{$B_2$\tablenotemark{b}}} &
\colhead{{$B_1 - B_2$}}
}
\tablenotetext{a}{The $F435W$ VEGAMAG ($B$-band equivalent) for the first $HST$ observation, taken at UT 21:35, 03-December-2003.}
\tablenotetext{b}{The $F435W$ VEGAMAG ($B$-band equivalent) for the second $HST$ observation, taken at UT 16:38, 01-March-2004.}
\startdata
J004357.5+411139 & 0:43:57.555 & 41:11:39.88 & 24.50$\pm$0.02 & 25.18$\pm$0.03 & -0.68$\pm$0.03\\
J004357.6+411119 & 0:43:57.659 & 41:11:19.99 & 24.90$\pm$0.02 & $>$27.8 & $<$-2.90\\
J004357.7+411126 & 0:43:57.722 & 41:11:26.35 & 25.20$\pm$0.03 & 27.10$\pm$0.14 & -1.90$\pm$0.15\\
J004357.9+411244 & 0:43:57.917 & 41:12:44.30 & 26.41$\pm$0.09 & 25.50$\pm$0.05 &  0.90$\pm$0.10\\
J004358.1+411250 & 0:43:58.194 & 41:12:50.15 & 23.22$\pm$0.01 & 23.88$\pm$0.01 & -0.66$\pm$0.01\\
J004358.5+411238 & 0:43:58.516 & 41:12:38.30 & 26.53$\pm$0.08 & 25.72$\pm$0.04 &  0.81$\pm$0.09\\
J004358.6+411109 & 0:43:58.607 & 41:11:09.19 & 25.42$\pm$0.03 & 26.81$\pm$0.12 & -1.39$\pm$0.12\\
J004358.6+411247 & 0:43:58.662 & 41:12:47.22 & 27.01$\pm$0.15 & 25.77$\pm$0.05 &  1.24$\pm$0.16\\
J004358.7+411211 & 0:43:58.701 & 41:12:11.11 & 24.60$\pm$0.02 & 25.41$\pm$0.03 & -0.81$\pm$0.04\\
J004359.0+411159 & 0:43:59.063 & 41:11:59.47 & 26.25$\pm$0.06 & 25.48$\pm$0.03 &  0.77$\pm$0.07\\
J004359.3+411125 & 0:43:59.368 & 41:11:25.85 & $>$27.8 & 23.76$\pm$0.01 & $>$4.04\\
J004359.5+411059 & 0:43:59.529 & 41:10:59.95 & $>$27.8 & 25.31$\pm$0.03 & $>$2.49\\
J004359.6+411221 & 0:43:59.613 & 41:12:21.77 & 24.33$\pm$0.01 & 27.70$\pm$0.20 & -3.38$\pm$0.20\\
J004359.8+411134 & 0:43:59.814 & 41:11:34.82 & 27.59$\pm$0.18 & 26.23$\pm$0.06 &  1.36$\pm$0.19\\
J004359.9+411126 & 0:43:59.964 & 41:11:26.22 & 25.93$\pm$0.05 & 25.08$\pm$0.02 &  0.85$\pm$0.05\\
J004400.0+411127 & 0:44:00.038 & 41:11:27.85 & 25.51$\pm$0.03 & 26.05$\pm$0.05 & -0.54$\pm$0.06\\
J004400.1+411323 & 0:44:00.167 & 41:13:23.35 & 25.14$\pm$0.03 & 25.94$\pm$0.05 & -0.80$\pm$0.05\\
J004400.3+411309 & 0:44:00.339 & 41:13:09.92 & 22.59$\pm$0.01 & 23.01$\pm$0.01 & -0.41$\pm$0.01\\
J004400.5+411215 & 0:44:00.524 & 41:12:15.24 & 25.03$\pm$0.02 & 24.59$\pm$0.02 &  0.44$\pm$0.03\\
J004400.6+411107 & 0:44:00.677 & 41:11:07.30 & 23.76$\pm$0.01 & 26.29$\pm$0.06 & -2.53$\pm$0.06\\
J004400.6+411239 & 0:44:00.683 & 41:12:39.68 & 25.61$\pm$0.03 & 25.37$\pm$0.03 &  0.24$\pm$0.04\\
J004400.6+411209 & 0:44:00.690 & 41:12:09.44 & 25.18$\pm$0.03 & $>$27.8 & $<$-2.62\\
J004400.7+411139 & 0:44:00.709 & 41:11:39.94 & 25.65$\pm$0.04 & 25.32$\pm$0.03 &  0.34$\pm$0.05\\
J004400.8+411243 & 0:44:00.804 & 41:12:43.30 & 25.03$\pm$0.03 & 25.35$\pm$0.03 & -0.31$\pm$0.04\\
J004400.8+411047 & 0:44:00.826 & 41:10:47.15 & 26.07$\pm$0.05 & 24.94$\pm$0.02 &  1.14$\pm$0.05\\
J004400.8+411305 & 0:44:00.842 & 41:13:05.86 & $>$27.8 & 27.00$\pm$0.12 & $>$0.80\\
J004400.9+411117 & 0:44:00.914 & 41:11:17.11 & 25.52$\pm$0.03 & $>$27.8 & $<$-2.28\\
J004400.9+411204 & 0:44:00.985 & 41:12:04.37 & 24.29$\pm$0.01 & $>$27.8 & $<$-3.51\\
J004400.9+411324 & 0:44:00.985 & 41:13:24.04 & 25.05$\pm$0.02 & 25.56$\pm$0.04 & -0.51$\pm$0.04\\
J004401.0+411342 & 0:44:01.025 & 41:13:42.44 & $>$27.8 & 24.60$\pm$0.02 & $>$3.20\\
J004401.0+411153 & 0:44:01.046 & 41:11:53.89 & 25.67$\pm$0.04 & 26.35$\pm$0.07 & -0.68$\pm$0.08\\
J004401.0+411132 & 0:44:01.084 & 41:11:32.35 & 22.13$\pm$0.01 & 23.06$\pm$0.01 & -0.93$\pm$0.01\\
J004401.2+411236 & 0:44:01.210 & 41:12:36.53 & 25.27$\pm$0.03 & 24.77$\pm$0.02 &  0.50$\pm$0.04\\
J004401.2+411233 & 0:44:01.241 & 41:12:33.02 & 24.90$\pm$0.02 & 25.57$\pm$0.04 & -0.67$\pm$0.04\\
J004401.2+411344 & 0:44:01.289 & 41:13:44.36 & 25.19$\pm$0.03 & 25.65$\pm$0.04 & -0.46$\pm$0.05\\
J004401.2+411042 & 0:44:01.294 & 41:10:42.94 & 25.20$\pm$0.02 & 27.11$\pm$0.12 & -1.91$\pm$0.12\\
J004401.4+411057 & 0:44:01.465 & 41:10:57.60 & 25.49$\pm$0.03 & 24.79$\pm$0.02 &  0.71$\pm$0.04\\
J004401.5+411333 & 0:44:01.536 & 41:13:33.40 & 25.30$\pm$0.03 & 25.97$\pm$0.05 & -0.66$\pm$0.06\\
J004401.5+411203 & 0:44:01.565 & 41:12:03.64 & 25.46$\pm$0.03 & 26.01$\pm$0.05 & -0.56$\pm$0.06\\
J004401.5+411203 & 0:44:01.580 & 41:12:03.23 & $>$27.8 & 26.11$\pm$0.05 & $>$1.69\\
J004401.6+411154 & 0:44:01.603 & 41:11:54.88 & 24.80$\pm$0.02 & 25.07$\pm$0.03 & -0.28$\pm$0.03\\
J004401.6+411327 & 0:44:01.663 & 41:13:27.35 & 25.66$\pm$0.04 & 26.41$\pm$0.07 & -0.75$\pm$0.08\\
J004401.8+411107 & 0:44:01.825 & 41:11:07.85 & 25.57$\pm$0.03 & 27.00$\pm$0.12 & -1.43$\pm$0.12\\
J004401.8+411133 & 0:44:01.861 & 41:11:33.61 & 26.39$\pm$0.07 & 25.75$\pm$0.05 &  0.63$\pm$0.09\\
J004402.0+411135 & 0:44:02.017 & 41:11:35.65 & 24.87$\pm$0.02 & 25.30$\pm$0.03 & -0.43$\pm$0.04\\
J004402.0+411347 & 0:44:02.074 & 41:13:47.94 & 24.66$\pm$0.02 & 24.95$\pm$0.02 & -0.29$\pm$0.03\\
J004402.0+411125 & 0:44:02.093 & 41:11:25.98 & 24.84$\pm$0.02 & 24.22$\pm$0.01 &  0.63$\pm$0.03\\
J004402.0+411215 & 0:44:02.093 & 41:12:15.05 & 25.22$\pm$0.03 & 25.05$\pm$0.03 &  0.18$\pm$0.04\\
J004402.1+411123 & 0:44:02.132 & 41:11:23.97 & 25.15$\pm$0.03 & 26.52$\pm$0.08 & -1.38$\pm$0.09\\
J004402.1+411233 & 0:44:02.162 & 41:12:33.64 & 25.07$\pm$0.03 & 24.83$\pm$0.02 &  0.24$\pm$0.03\\
J004402.2+411342 & 0:44:02.282 & 41:13:42.00 & 25.22$\pm$0.03 & 26.39$\pm$0.08 & -1.16$\pm$0.08\\
J004402.2+411346 & 0:44:02.288 & 41:13:46.06 & 24.90$\pm$0.02 & 25.21$\pm$0.03 & -0.30$\pm$0.04\\
J004402.5+411137 & 0:44:02.512 & 41:11:37.10 & 22.57$\pm$0.01 & 22.09$\pm$0.00 &  0.48$\pm$0.01\\
J004402.5+411121 & 0:44:02.564 & 41:11:21.89 & 26.46$\pm$0.07 & 25.81$\pm$0.04 &  0.65$\pm$0.08\\
J004402.6+411214 & 0:44:02.634 & 41:12:14.79 & 24.93$\pm$0.02 & 25.76$\pm$0.04 & -0.84$\pm$0.05\\
J004402.6+411140 & 0:44:02.642 & 41:11:40.45 & 26.26$\pm$0.08 & 25.43$\pm$0.04 &  0.83$\pm$0.09\\
J004402.6+411148 & 0:44:02.650 & 41:11:48.66 & 25.32$\pm$0.03 & 26.39$\pm$0.07 & -1.06$\pm$0.08\\
J004402.7+411348 & 0:44:02.779 & 41:13:48.34 & 25.55$\pm$0.05 & 25.22$\pm$0.04 &  0.34$\pm$0.06\\
J004402.9+411224 & 0:44:02.928 & 41:12:24.95 & 25.52$\pm$0.04 & 26.45$\pm$0.07 & -0.94$\pm$0.08\\
J004402.9+411320 & 0:44:02.973 & 41:13:20.86 & 25.83$\pm$0.04 & 26.37$\pm$0.07 & -0.54$\pm$0.09\\
J004403.0+411233 & 0:44:03.096 & 41:12:33.59 & 25.52$\pm$0.03 & 26.62$\pm$0.09 & -1.10$\pm$0.09\\
J004403.2+411137 & 0:44:03.221 & 41:11:37.71 & 24.89$\pm$0.02 & 26.80$\pm$0.10 & -1.91$\pm$0.10\\
J004403.3+411047 & 0:44:03.318 & 41:10:47.98 & 25.04$\pm$0.02 & 24.76$\pm$0.02 &  0.28$\pm$0.03\\
J004403.3+411050 & 0:44:03.343 & 41:10:50.37 & 25.79$\pm$0.05 & 24.83$\pm$0.03 &  0.96$\pm$0.05\\
J004403.3+411325 & 0:44:03.341 & 41:13:25.15 & 25.38$\pm$0.03 & 26.64$\pm$0.08 & -1.26$\pm$0.09\\
J004403.4+411328 & 0:44:03.417 & 41:13:28.24 & 25.20$\pm$0.02 & 24.78$\pm$0.02 &  0.42$\pm$0.03\\
J004403.4+411103 & 0:44:03.435 & 41:11:03.70 & 25.75$\pm$0.04 & 27.55$\pm$0.19 & -1.80$\pm$0.19\\
J004403.4+411307 & 0:44:03.463 & 41:13:07.80 & 26.35$\pm$0.07 & 25.60$\pm$0.04 &  0.75$\pm$0.08\\
J004403.5+411228 & 0:44:03.558 & 41:12:28.26 & $>$27.8 & 26.77$\pm$0.10 & $>$1.03\\
J004403.6+411117 & 0:44:03.633 & 41:11:17.31 & 26.65$\pm$0.08 & 25.80$\pm$0.04 &  0.85$\pm$0.09\\
J004403.7+411339 & 0:44:03.753 & 41:13:39.11 & 24.98$\pm$0.03 & 25.75$\pm$0.04 & -0.76$\pm$0.05\\
J004403.7+411250 & 0:44:03.798 & 41:12:50.84 & 25.59$\pm$0.04 & 26.80$\pm$0.11 & -1.21$\pm$0.11\\
J004403.8+411137 & 0:44:03.811 & 41:11:37.88 & 25.55$\pm$0.04 & 25.09$\pm$0.03 &  0.46$\pm$0.05\\
J004403.8+411323 & 0:44:03.830 & 41:13:23.03 & 25.55$\pm$0.04 & 26.45$\pm$0.08 & -0.90$\pm$0.08\\
J004403.8+411129 & 0:44:03.864 & 41:11:29.63 & 24.70$\pm$0.02 & 25.21$\pm$0.03 & -0.51$\pm$0.03\\
J004403.9+411357 & 0:44:03.942 & 41:13:57.52 & 26.76$\pm$0.11 & 24.71$\pm$0.02 &  2.06$\pm$0.11\\
J004403.9+411143 & 0:44:03.982 & 41:11:43.87 & 24.40$\pm$0.01 & 25.11$\pm$0.03 & -0.71$\pm$0.03\\
J004404.0+411353 & 0:44:04.030 & 41:13:53.66 & 22.01$\pm$0.00 & 22.97$\pm$0.01 & -0.96$\pm$0.01\\
J004404.1+411243 & 0:44:04.145 & 41:12:43.12 & 25.59$\pm$0.04 & 26.67$\pm$0.09 & -1.09$\pm$0.09\\
J004404.1+411047 & 0:44:04.154 & 41:10:47.86 & 26.32$\pm$0.07 & 25.39$\pm$0.03 &  0.93$\pm$0.07\\
J004404.1+411111 & 0:44:04.159 & 41:11:11.29 & 26.67$\pm$0.08 & 26.04$\pm$0.05 &  0.63$\pm$0.09\\
J004404.1+411152 & 0:44:04.188 & 41:11:52.39 & 26.47$\pm$0.08 & 25.61$\pm$0.04 &  0.86$\pm$0.09\\
J004404.1+411250 & 0:44:04.197 & 41:12:50.66 & 26.40$\pm$0.06 & 25.84$\pm$0.04 &  0.56$\pm$0.07\\
J004404.2+411303 & 0:44:04.235 & 41:13:03.54 & 26.32$\pm$0.07 & 25.52$\pm$0.03 &  0.79$\pm$0.07\\
J004404.2+411235 & 0:44:04.275 & 41:12:35.73 & 25.24$\pm$0.03 & 26.42$\pm$0.07 & -1.18$\pm$0.08\\
J004404.2+411213 & 0:44:04.286 & 41:12:13.63 & 25.30$\pm$0.03 & 25.63$\pm$0.04 & -0.34$\pm$0.05\\
J004404.3+411328 & 0:44:04.319 & 41:13:28.71 & 25.33$\pm$0.03 & 24.96$\pm$0.02 &  0.38$\pm$0.04\\
J004404.4+411129 & 0:44:04.402 & 41:11:29.44 & 24.82$\pm$0.02 & 26.29$\pm$0.06 & -1.47$\pm$0.07\\
J004404.4+411230 & 0:44:04.480 & 41:12:30.24 & 27.17$\pm$0.11 & 25.86$\pm$0.04 &  1.31$\pm$0.12\\
J004404.6+411232 & 0:44:04.623 & 41:12:32.38 & 26.52$\pm$0.08 & 23.87$\pm$0.01 &  2.65$\pm$0.08\\
J004404.6+411100 & 0:44:04.684 & 41:11:00.38 & 27.48$\pm$0.14 & 23.62$\pm$0.01 &  3.86$\pm$0.14\\
J004404.6+411331 & 0:44:04.687 & 41:13:31.97 & 26.24$\pm$0.06 & 25.50$\pm$0.04 &  0.74$\pm$0.07\\
J004404.6+411339 & 0:44:04.694 & 41:13:39.76 & 25.44$\pm$0.03 & 25.87$\pm$0.04 & -0.43$\pm$0.05\\
J004404.7+411124 & 0:44:04.714 & 41:11:24.97 & 26.33$\pm$0.07 & 25.49$\pm$0.04 &  0.83$\pm$0.08\\
J004404.7+411343 & 0:44:04.725 & 41:13:43.86 & 25.40$\pm$0.03 & 26.47$\pm$0.07 & -1.07$\pm$0.07\\
J004404.7+411202 & 0:44:04.778 & 41:12:02.94 & $>$27.8 & 25.21$\pm$0.03 & $>$2.59\\
J004404.8+411136 & 0:44:04.808 & 41:11:36.46 & 25.38$\pm$0.04 & 24.89$\pm$0.03 &  0.49$\pm$0.04\\
J004404.9+411130 & 0:44:04.916 & 41:11:30.51 & 25.41$\pm$0.03 & 25.85$\pm$0.04 & -0.43$\pm$0.05\\
J004404.9+411239 & 0:44:04.917 & 41:12:39.40 & 24.99$\pm$0.02 & 25.79$\pm$0.04 & -0.79$\pm$0.05\\
J004404.9+411235 & 0:44:04.930 & 41:12:35.18 & 25.27$\pm$0.03 & 25.77$\pm$0.04 & -0.50$\pm$0.05\\
J004404.9+411321 & 0:44:04.946 & 41:13:21.48 & 24.35$\pm$0.01 & $>$27.8 & $<$-3.45\\
J004404.9+411351 & 0:44:04.977 & 41:13:51.91 & 25.03$\pm$0.02 & 25.26$\pm$0.03 & -0.23$\pm$0.04\\
J004404.9+411228 & 0:44:04.991 & 41:12:28.37 & 24.32$\pm$0.01 & 24.67$\pm$0.02 & -0.35$\pm$0.02\\
J004405.1+411329 & 0:44:05.134 & 41:13:29.60 & 27.40$\pm$0.17 & 23.95$\pm$0.01 &  3.44$\pm$0.17\\
J004405.1+411158 & 0:44:05.149 & 41:11:58.63 & 25.38$\pm$0.03 & 25.15$\pm$0.03 &  0.24$\pm$0.04\\
J004405.2+411221 & 0:44:05.205 & 41:12:21.37 & 25.71$\pm$0.04 & 25.11$\pm$0.02 &  0.60$\pm$0.04\\
J004405.2+411341 & 0:44:05.208 & 41:13:41.48 & 26.33$\pm$0.08 & 25.48$\pm$0.04 &  0.85$\pm$0.09\\
J004405.2+411318 & 0:44:05.272 & 41:13:18.52 & 26.03$\pm$0.06 & 25.40$\pm$0.03 &  0.63$\pm$0.06\\
J004405.4+411245 & 0:44:05.472 & 41:12:45.25 & 25.35$\pm$0.03 & 26.07$\pm$0.06 & -0.73$\pm$0.06\\
J004405.7+411127 & 0:44:05.756 & 41:11:27.86 & 25.45$\pm$0.03 & 26.25$\pm$0.07 & -0.80$\pm$0.08\\
J004405.7+411049 & 0:44:05.780 & 41:10:49.95 & 24.82$\pm$0.02 & 25.77$\pm$0.04 & -0.95$\pm$0.05\\
J004405.7+411202 & 0:44:05.780 & 41:12:02.58 & 24.64$\pm$0.02 & 24.13$\pm$0.01 &  0.51$\pm$0.02\\
J004405.8+411049 & 0:44:05.893 & 41:10:49.84 & 25.51$\pm$0.03 & 25.78$\pm$0.04 & -0.27$\pm$0.05\\
J004405.9+411316 & 0:44:05.902 & 41:13:16.97 & 26.77$\pm$0.08 & 25.64$\pm$0.03 &  1.12$\pm$0.08\\
J004406.0+411348 & 0:44:06.064 & 41:13:48.28 & $>$27.8 & 24.50$\pm$0.02 & $>$3.30\\
J004406.0+411049 & 0:44:06.073 & 41:10:49.97 & 25.81$\pm$0.04 & 26.20$\pm$0.06 & -0.38$\pm$0.08\\
J004406.1+411235 & 0:44:06.120 & 41:12:35.04 & 23.52$\pm$0.01 & 23.91$\pm$0.01 & -0.38$\pm$0.02\\
J004406.1+411105 & 0:44:06.199 & 41:11:05.71 & 25.50$\pm$0.03 & 25.24$\pm$0.03 &  0.27$\pm$0.04\\
J004406.2+411354 & 0:44:06.256 & 41:13:54.00 & 25.54$\pm$0.03 & 26.65$\pm$0.08 & -1.12$\pm$0.09\\
J004406.2+411052 & 0:44:06.277 & 41:10:52.21 & 26.18$\pm$0.05 & 25.62$\pm$0.04 &  0.56$\pm$0.06\\
J004406.3+411305 & 0:44:06.301 & 41:13:05.49 & 26.84$\pm$0.10 & 25.95$\pm$0.05 &  0.88$\pm$0.11\\
J004406.3+411127 & 0:44:06.307 & 41:11:27.99 & 24.65$\pm$0.02 & 25.14$\pm$0.03 & -0.49$\pm$0.03\\
J004406.3+411223 & 0:44:06.335 & 41:12:23.97 & 25.55$\pm$0.03 & 25.21$\pm$0.03 &  0.34$\pm$0.04\\
J004406.3+411346 & 0:44:06.367 & 41:13:46.42 & 24.75$\pm$0.02 & 25.13$\pm$0.03 & -0.38$\pm$0.03\\
J004406.3+411238 & 0:44:06.395 & 41:12:38.33 & 24.90$\pm$0.02 & 26.02$\pm$0.05 & -1.12$\pm$0.06\\
J004406.4+411340 & 0:44:06.411 & 41:13:40.76 & 24.50$\pm$0.02 & 25.83$\pm$0.04 & -1.33$\pm$0.04\\
J004406.4+411234 & 0:44:06.418 & 41:12:34.26 & 26.89$\pm$0.11 & $>$27.8 & $<$-0.91\\
J004406.4+411404 & 0:44:06.485 & 41:14:04.69 & 26.00$\pm$0.05 & 25.62$\pm$0.04 &  0.38$\pm$0.06\\
J004406.5+411200 & 0:44:06.545 & 41:12:00.76 & 26.32$\pm$0.07 & 25.94$\pm$0.05 &  0.38$\pm$0.08\\
J004406.5+411158 & 0:44:06.573 & 41:11:58.46 & 24.13$\pm$0.01 & 27.25$\pm$0.16 & -3.12$\pm$0.16\\
J004406.6+411122 & 0:44:06.606 & 41:11:22.28 & 25.23$\pm$0.03 & 24.42$\pm$0.02 &  0.81$\pm$0.03\\
J004406.6+411346 & 0:44:06.647 & 41:13:46.98 & 26.55$\pm$0.07 & 25.85$\pm$0.05 &  0.71$\pm$0.09\\
J004406.6+411342 & 0:44:06.655 & 41:13:42.66 & $>$27.8 & 24.75$\pm$0.02 & $>$3.05\\
J004406.7+411323 & 0:44:06.736 & 41:13:23.15 & 26.06$\pm$0.05 & 25.29$\pm$0.03 &  0.77$\pm$0.06\\
J004406.8+411325 & 0:44:06.805 & 41:13:25.34 & 24.51$\pm$0.02 & 24.33$\pm$0.01 &  0.18$\pm$0.02\\
J004406.8+411224 & 0:44:06.807 & 41:12:24.03 & 25.85$\pm$0.04 & 26.82$\pm$0.09 & -0.98$\pm$0.10\\
J004406.9+411340 & 0:44:06.918 & 41:13:40.04 & 25.65$\pm$0.04 & 26.64$\pm$0.09 & -0.99$\pm$0.10\\
J004407.0+411254 & 0:44:07.022 & 41:12:54.48 & 24.15$\pm$0.01 & 25.05$\pm$0.03 & -0.90$\pm$0.03\\
J004407.0+411146 & 0:44:07.034 & 41:11:46.82 & 24.38$\pm$0.01 & 24.21$\pm$0.01 &  0.17$\pm$0.02\\
J004407.0+411047 & 0:44:07.067 & 41:10:47.99 & 26.70$\pm$0.09 & 25.44$\pm$0.03 &  1.25$\pm$0.09\\
J004407.2+411047 & 0:44:07.202 & 41:10:47.64 & 26.14$\pm$0.05 & 25.58$\pm$0.04 &  0.56$\pm$0.06\\
J004407.2+411354 & 0:44:07.235 & 41:13:54.65 & 25.42$\pm$0.04 & 26.68$\pm$0.11 & -1.26$\pm$0.12\\
J004407.2+411226 & 0:44:07.297 & 41:12:26.00 & 25.17$\pm$0.03 & 24.92$\pm$0.02 &  0.24$\pm$0.03\\
J004407.3+411300 & 0:44:07.395 & 41:13:00.72 & 25.89$\pm$0.04 & 25.40$\pm$0.03 &  0.49$\pm$0.05\\
J004407.4+411119 & 0:44:07.400 & 41:11:19.66 & 24.78$\pm$0.02 & 27.18$\pm$0.13 & -2.40$\pm$0.13\\
J004407.4+411204 & 0:44:07.414 & 41:12:04.85 & 24.57$\pm$0.02 & 25.25$\pm$0.03 & -0.68$\pm$0.04\\
J004407.5+411321 & 0:44:07.506 & 41:13:21.46 & 25.77$\pm$0.04 & 25.32$\pm$0.03 &  0.45$\pm$0.05\\
J004407.5+411350 & 0:44:07.533 & 41:13:50.34 & 25.59$\pm$0.03 & 25.08$\pm$0.02 &  0.50$\pm$0.04\\
J004407.5+411126 & 0:44:07.561 & 41:11:26.59 & 25.54$\pm$0.03 & 26.28$\pm$0.06 & -0.74$\pm$0.06\\
J004407.5+411304 & 0:44:07.584 & 41:13:04.50 & 25.24$\pm$0.03 & 25.78$\pm$0.05 & -0.54$\pm$0.05\\
J004407.6+411232 & 0:44:07.625 & 41:12:32.91 & 26.06$\pm$0.06 & 25.55$\pm$0.04 &  0.51$\pm$0.07\\
J004407.6+411238 & 0:44:07.692 & 41:12:38.51 & 25.69$\pm$0.04 & 25.30$\pm$0.03 &  0.40$\pm$0.05\\
J004407.7+411110 & 0:44:07.721 & 41:11:10.04 & 26.18$\pm$0.05 & 25.34$\pm$0.03 &  0.84$\pm$0.06\\
J004407.7+411123 & 0:44:07.733 & 41:11:23.74 & 25.29$\pm$0.03 & 26.04$\pm$0.05 & -0.75$\pm$0.06\\
J004407.7+411357 & 0:44:07.764 & 41:13:57.65 & 24.94$\pm$0.02 & 23.22$\pm$0.01 &  1.72$\pm$0.02\\
J004407.7+411331 & 0:44:07.768 & 41:13:31.48 & $>$27.8 & 24.58$\pm$0.02 & $>$3.22\\
J004407.7+411333 & 0:44:07.798 & 41:13:33.38 & 25.26$\pm$0.03 & 26.31$\pm$0.07 & -1.04$\pm$0.07\\
J004407.8+411208 & 0:44:07.822 & 41:12:08.62 & 23.43$\pm$0.01 & 25.58$\pm$0.04 & -2.15$\pm$0.04\\
J004407.9+411327 & 0:44:07.930 & 41:13:27.64 & 25.10$\pm$0.02 & 24.87$\pm$0.02 &  0.23$\pm$0.03\\
J004408.0+411145 & 0:44:08.017 & 41:11:45.58 & 24.02$\pm$0.01 & 25.36$\pm$0.03 & -1.34$\pm$0.03\\
J004408.0+411243 & 0:44:08.049 & 41:12:43.41 & 23.28$\pm$0.01 & 22.39$\pm$0.01 &  0.89$\pm$0.01\\
J004408.1+411058 & 0:44:08.103 & 41:10:58.51 & 26.94$\pm$0.09 & 26.37$\pm$0.06 &  0.57$\pm$0.11\\
J004408.1+411302 & 0:44:08.178 & 41:13:02.30 & 25.35$\pm$0.03 & 25.98$\pm$0.05 & -0.62$\pm$0.06\\
J004408.2+411212 & 0:44:08.229 & 41:12:12.03 & $>$27.8 & 25.11$\pm$0.03 & $>$2.69\\
J004408.3+411202 & 0:44:08.313 & 41:12:02.93 & 26.60$\pm$0.09 & 25.55$\pm$0.04 &  1.04$\pm$0.10\\
J004408.4+411239 & 0:44:08.438 & 41:12:39.55 & 24.80$\pm$0.02 & 24.96$\pm$0.03 & -0.16$\pm$0.03\\
J004408.4+411347 & 0:44:08.492 & 41:13:47.30 & 21.92$\pm$0.00 & 22.50$\pm$0.01 & -0.58$\pm$0.01\\
J004408.5+411156 & 0:44:08.547 & 41:11:56.42 & 25.51$\pm$0.03 & 26.68$\pm$0.10 & -1.17$\pm$0.10\\
J004408.5+411244 & 0:44:08.597 & 41:12:44.95 & 25.12$\pm$0.03 & 24.68$\pm$0.02 &  0.44$\pm$0.03\\
J004408.6+411229 & 0:44:08.617 & 41:12:29.70 & 26.59$\pm$0.10 & 24.66$\pm$0.02 &  1.93$\pm$0.10\\
J004408.6+411236 & 0:44:08.625 & 41:12:36.45 & 25.94$\pm$0.05 & 25.56$\pm$0.04 &  0.37$\pm$0.06\\
J004408.6+411331 & 0:44:08.653 & 41:13:31.35 & 25.67$\pm$0.05 & $>$27.8 & $<$-2.13\\
J004408.6+411106 & 0:44:08.699 & 41:11:06.83 & 25.78$\pm$0.04 & 25.45$\pm$0.03 &  0.33$\pm$0.05\\
J004408.7+411129 & 0:44:08.738 & 41:11:29.87 & 23.76$\pm$0.01 & 24.33$\pm$0.01 & -0.57$\pm$0.02\\
J004408.7+411319 & 0:44:08.744 & 41:13:19.36 & 25.33$\pm$0.03 & 24.98$\pm$0.03 &  0.36$\pm$0.04\\
J004408.8+411330 & 0:44:08.830 & 41:13:30.95 & 26.03$\pm$0.05 & 27.13$\pm$0.12 & -1.11$\pm$0.13\\
J004408.8+411235 & 0:44:08.867 & 41:12:35.37 & 27.58$\pm$0.17 & 26.18$\pm$0.05 &  1.40$\pm$0.18\\
J004408.9+411101 & 0:44:08.907 & 41:11:01.58 & 26.21$\pm$0.06 & 25.31$\pm$0.03 &  0.89$\pm$0.06\\
J004408.9+411134 & 0:44:08.945 & 41:11:34.73 & 25.05$\pm$0.03 & 26.34$\pm$0.07 & -1.29$\pm$0.08\\
J004408.9+411051 & 0:44:08.953 & 41:10:51.74 & 25.54$\pm$0.03 & 25.24$\pm$0.02 &  0.30$\pm$0.04\\
J004409.0+411316 & 0:44:09.040 & 41:13:16.80 & 25.49$\pm$0.04 & 26.51$\pm$0.08 & -1.02$\pm$0.09\\
J004409.0+411228 & 0:44:09.094 & 41:12:28.96 & 25.00$\pm$0.02 & 25.44$\pm$0.03 & -0.44$\pm$0.04\\
J004409.1+411119 & 0:44:09.123 & 41:11:19.56 & 24.98$\pm$0.02 & 24.75$\pm$0.02 &  0.23$\pm$0.03\\
J004409.3+411339 & 0:44:09.307 & 41:13:39.94 & 25.63$\pm$0.04 & 26.27$\pm$0.06 & -0.65$\pm$0.07\\
J004409.3+411226 & 0:44:09.347 & 41:12:26.45 & 25.19$\pm$0.03 & $>$27.8 & $<$-2.61\\
J004409.3+411202 & 0:44:09.381 & 41:12:02.61 & 24.58$\pm$0.02 & 26.11$\pm$0.06 & -1.53$\pm$0.06\\
J004409.3+411325 & 0:44:09.393 & 41:13:25.09 & 25.58$\pm$0.03 & 26.33$\pm$0.06 & -0.75$\pm$0.07\\
J004409.4+411311 & 0:44:09.407 & 41:13:11.84 & $>$27.8 & 26.00$\pm$0.05 & $>$1.80\\
J004409.4+411203 & 0:44:09.418 & 41:12:03.53 & 23.99$\pm$0.01 & 26.22$\pm$0.06 & -2.24$\pm$0.06\\
J004409.4+411247 & 0:44:09.420 & 41:12:47.29 & 24.64$\pm$0.02 & 24.25$\pm$0.02 &  0.38$\pm$0.02\\
J004409.4+411207 & 0:44:09.446 & 41:12:07.94 & 24.68$\pm$0.02 & 26.44$\pm$0.08 & -1.76$\pm$0.08\\
J004409.5+411147 & 0:44:09.560 & 41:11:47.25 & 26.40$\pm$0.06 & 25.64$\pm$0.04 &  0.76$\pm$0.08\\
J004409.6+411325 & 0:44:09.682 & 41:13:25.82 & 25.94$\pm$0.05 & 26.87$\pm$0.12 & -0.93$\pm$0.13\\
J004409.7+411243 & 0:44:09.713 & 41:12:43.76 & 24.02$\pm$0.01 & 26.43$\pm$0.07 & -2.41$\pm$0.07\\
J004409.7+411052 & 0:44:09.728 & 41:10:52.24 & 26.25$\pm$0.06 & 25.03$\pm$0.03 &  1.22$\pm$0.07\\
J004409.7+411115 & 0:44:09.735 & 41:11:15.05 & 25.58$\pm$0.03 & 25.21$\pm$0.03 &  0.37$\pm$0.04\\
J004409.8+411256 & 0:44:09.832 & 41:12:56.83 & 24.28$\pm$0.01 & 26.72$\pm$0.09 & -2.45$\pm$0.09\\
J004409.9+411108 & 0:44:09.901 & 41:11:08.48 & 26.34$\pm$0.06 & 25.72$\pm$0.04 &  0.62$\pm$0.07\\
J004409.9+411201 & 0:44:09.910 & 41:12:01.24 & 24.58$\pm$0.02 & 26.67$\pm$0.09 & -2.09$\pm$0.09\\
J004410.0+411213 & 0:44:10.039 & 41:12:13.15 & 26.36$\pm$0.06 & 25.73$\pm$0.04 &  0.63$\pm$0.07\\
J004410.1+411244 & 0:44:10.182 & 41:12:44.20 & 25.64$\pm$0.04 & 25.12$\pm$0.03 &  0.52$\pm$0.04\\
J004410.2+411229 & 0:44:10.212 & 41:12:29.35 & 25.94$\pm$0.05 & 24.92$\pm$0.02 &  1.02$\pm$0.05\\
J004410.2+411258 & 0:44:10.237 & 41:12:58.67 & 26.98$\pm$0.12 & 25.40$\pm$0.03 &  1.58$\pm$0.12\\
J004410.3+411321 & 0:44:10.309 & 41:13:21.70 & 27.79$\pm$0.23 & 23.63$\pm$0.01 &  4.16$\pm$0.23\\
J004410.3+411146 & 0:44:10.332 & 41:11:46.91 & 25.72$\pm$0.04 & 25.17$\pm$0.03 &  0.56$\pm$0.04\\
J004410.3+411052 & 0:44:10.381 & 41:10:52.82 & 26.11$\pm$0.05 & 27.07$\pm$0.11 & -0.96$\pm$0.12\\
J004410.4+411108 & 0:44:10.444 & 41:11:08.31 & 24.42$\pm$0.01 & 26.01$\pm$0.04 & -1.59$\pm$0.05\\
J004410.5+411053 & 0:44:10.503 & 41:10:53.49 & 26.00$\pm$0.05 & 25.54$\pm$0.03 &  0.46$\pm$0.06\\
J004410.6+411335 & 0:44:10.696 & 41:13:35.66 & 24.43$\pm$0.02 & 25.96$\pm$0.05 & -1.54$\pm$0.05\\
J004410.7+411110 & 0:44:10.714 & 41:11:10.27 & 25.40$\pm$0.03 & 26.81$\pm$0.10 & -1.41$\pm$0.10\\
J004410.9+411107 & 0:44:10.908 & 41:11:07.66 & 25.08$\pm$0.03 & 25.86$\pm$0.05 & -0.78$\pm$0.05\\
J004410.9+411132 & 0:44:10.915 & 41:11:32.53 & 25.94$\pm$0.04 & 25.56$\pm$0.03 &  0.37$\pm$0.05\\
J004410.9+411104 & 0:44:10.935 & 41:11:04.07 & 24.58$\pm$0.02 & 24.96$\pm$0.02 & -0.38$\pm$0.03\\
J004411.1+411115 & 0:44:11.152 & 41:11:15.72 & 26.35$\pm$0.06 & 25.76$\pm$0.04 &  0.59$\pm$0.07\\
J004411.1+411228 & 0:44:11.159 & 41:12:28.96 & 25.98$\pm$0.04 & 25.31$\pm$0.03 &  0.66$\pm$0.05\\
J004411.1+411102 & 0:44:11.182 & 41:11:02.57 & 26.77$\pm$0.08 & 26.03$\pm$0.04 &  0.74$\pm$0.09\\
J004411.1+411154 & 0:44:11.188 & 41:11:54.73 & 25.25$\pm$0.03 & 25.09$\pm$0.03 &  0.16$\pm$0.04\\
J004411.2+411056 & 0:44:11.250 & 41:10:56.43 & 26.52$\pm$0.08 & $>$27.8 & $<$-1.28\\
J004411.2+411102 & 0:44:11.254 & 41:11:02.99 & 24.98$\pm$0.02 & 27.48$\pm$0.17 & -2.50$\pm$0.17\\
J004411.3+411129 & 0:44:11.397 & 41:11:29.75 & 25.19$\pm$0.03 & 25.85$\pm$0.04 & -0.66$\pm$0.05\\
J004411.4+411116 & 0:44:11.434 & 41:11:16.91 & 25.86$\pm$0.04 & 25.58$\pm$0.03 &  0.28$\pm$0.05\\
J004411.4+411111 & 0:44:11.446 & 41:11:11.97 & 25.61$\pm$0.03 & 25.97$\pm$0.05 & -0.36$\pm$0.06\\
J004411.5+411154 & 0:44:11.546 & 41:11:54.28 & 27.49$\pm$0.14 & 26.49$\pm$0.07 &  0.99$\pm$0.15\\
J004411.6+411140 & 0:44:11.639 & 41:11:40.51 & 24.64$\pm$0.02 & 26.35$\pm$0.07 & -1.71$\pm$0.07\\
J004411.6+411228 & 0:44:11.669 & 41:12:28.22 & 25.24$\pm$0.03 & 26.53$\pm$0.08 & -1.29$\pm$0.08\\
J004411.8+411107 & 0:44:11.817 & 41:11:07.49 & 25.74$\pm$0.04 & $>$27.8 & $<$-2.06\\
J004411.8+411058 & 0:44:11.822 & 41:10:58.62 & 25.47$\pm$0.03 & 25.97$\pm$0.05 & -0.51$\pm$0.06\\
J004412.0+411204 & 0:44:12.084 & 41:12:04.06 & 24.60$\pm$0.02 & 24.98$\pm$0.02 & -0.38$\pm$0.03\\
J004412.0+411151 & 0:44:12.092 & 41:11:51.38 & 26.17$\pm$0.05 & 25.73$\pm$0.04 &  0.44$\pm$0.07\\
J004412.1+411137 & 0:44:12.108 & 41:11:37.86 & 27.29$\pm$0.12 & 26.05$\pm$0.05 &  1.24$\pm$0.13\\
J004412.2+411112 & 0:44:12.219 & 41:11:12.14 & 25.63$\pm$0.04 & 25.06$\pm$0.02 &  0.57$\pm$0.04\\
J004412.2+411153 & 0:44:12.263 & 41:11:53.71 & 24.84$\pm$0.02 & 25.07$\pm$0.02 & -0.23$\pm$0.03\\
J004412.3+411156 & 0:44:12.343 & 41:11:56.25 & 26.63$\pm$0.08 & 26.09$\pm$0.05 &  0.54$\pm$0.10\\
J004412.3+411121 & 0:44:12.344 & 41:11:21.19 & 24.87$\pm$0.02 & $>$27.8 & $<$-2.93\\
J004412.4+411131 & 0:44:12.405 & 41:11:31.72 & 25.74$\pm$0.04 & 26.05$\pm$0.05 & -0.30$\pm$0.06\\
J004412.4+411318 & 0:44:12.415 & 41:13:18.76 & 24.74$\pm$0.02 & 25.48$\pm$0.03 & -0.74$\pm$0.04\\
J004412.4+411229 & 0:44:12.455 & 41:12:29.15 & $>$27.8 & 24.64$\pm$0.02 & $>$3.16\\
J004412.5+411058 & 0:44:12.510 & 41:10:58.69 & 24.77$\pm$0.02 & $>$27.8 & $<$-3.03\\
J004412.9+411231 & 0:44:12.912 & 41:12:31.43 & 24.99$\pm$0.02 & 25.60$\pm$0.04 & -0.61$\pm$0.04\\
J004412.9+411153 & 0:44:12.957 & 41:11:53.58 & 25.21$\pm$0.03 & 25.52$\pm$0.03 & -0.30$\pm$0.04\\
J004412.9+411152 & 0:44:12.998 & 41:11:52.77 & 25.47$\pm$0.03 & 25.85$\pm$0.05 & -0.38$\pm$0.06\\
J004413.1+411152 & 0:44:13.166 & 41:11:52.45 & 24.98$\pm$0.02 & 25.53$\pm$0.03 & -0.55$\pm$0.04\\
J004413.2+411139 & 0:44:13.264 & 41:11:39.21 & 25.28$\pm$0.03 & 26.21$\pm$0.06 & -0.93$\pm$0.07\\
J004413.3+411308 & 0:44:13.317 & 41:13:08.24 & 24.88$\pm$0.02 & 23.90$\pm$0.01 &  0.98$\pm$0.02\\
J004413.3+411142 & 0:44:13.374 & 41:11:42.08 & 25.94$\pm$0.04 & 27.28$\pm$0.15 & -1.34$\pm$0.16\\
J004413.5+411204 & 0:44:13.514 & 41:12:04.22 & 24.87$\pm$0.02 & $>$27.8 & $<$-2.93\\
J004413.5+411216 & 0:44:13.527 & 41:12:16.09 & 24.80$\pm$0.02 & 25.27$\pm$0.03 & -0.47$\pm$0.03\\
J004413.6+411159 & 0:44:13.689 & 41:11:59.63 & 26.00$\pm$0.04 & $>$27.8 & $<$-1.80\\
J004413.8+411143 & 0:44:13.839 & 41:11:43.07 & 24.57$\pm$0.02 & 24.88$\pm$0.02 & -0.31$\pm$0.03\\
J004414.0+411155 & 0:44:14.091 & 41:11:55.09 & 25.08$\pm$0.02 & 26.64$\pm$0.08 & -1.56$\pm$0.08\\
J004414.2+411149 & 0:44:14.219 & 41:11:49.63 & 25.97$\pm$0.04 & 25.37$\pm$0.03 &  0.60$\pm$0.05\\
J004414.3+411226 & 0:44:14.325 & 41:12:26.56 & 26.30$\pm$0.07 & 25.50$\pm$0.03 &  0.80$\pm$0.08\\
J004414.3+411218 & 0:44:14.336 & 41:12:18.20 & 23.87$\pm$0.01 & $>$27.8 & $<$-3.93\\
J004415.2+411157 & 0:44:15.209 & 41:11:57.32 & 26.17$\pm$0.06 & 24.40$\pm$0.02 &  1.77$\pm$0.06\\
\enddata
\tablecomments{The complete version of this table is in the electronic
edition of the Journal.  The printed edition contains only a sample.}
\label{cat}
\end{deluxetable}

\begin{deluxetable}{cccc}
\tablecaption{Number of variable stars in the catalog and their photometric properties in each 0.5 magnitude bin.}
\tableheadfrac{0.01}
\tablehead{
\colhead{{$B_{bright}$}\tablenotemark{a}} &
\colhead{{Number}\tablenotemark{b}} &
\colhead{{Mean S/N ($\sigma$)\tablenotemark{c}}} &
\colhead{{Mean $B_{err}$}\tablenotemark{d}}
}
\tablenotetext{a}{The range of magnitudes included in the bin.  Only the brighter of the two epochs is included.}
\tablenotetext{b}{The number of variable candidates found whose brighter detection was within the indicated magnitude range.}
\tablenotetext{c}{The mean significance of the detections in this magnitude range in $\sigma$.}
\tablenotetext{d}{The mean photometric errors of the detections in this magnitude range in magnitudes.}
\startdata
21.50 $\rightarrow$ 22.00 &   1 & 268 & 0.004\\
22.00 $\rightarrow$ 22.50 &   4 & 239 & 0.005\\
22.50 $\rightarrow$ 23.00 &   1 & 193 & 0.006\\
23.00 $\rightarrow$ 23.50 &   3 & 134 & 0.008\\
23.50 $\rightarrow$ 24.00 &  11 & 102 & 0.011\\
24.00 $\rightarrow$ 24.50 &  22 &  76 & 0.015\\
24.50 $\rightarrow$ 25.00 &  58 &  56 & 0.020\\
25.00 $\rightarrow$ 25.50 &  81 &  40 & 0.029\\
25.50 $\rightarrow$ 26.00 &  57 &  30 & 0.039\\
26.00 $\rightarrow$ 26.50 &  12 &  20 & 0.052\\
26.50 $\rightarrow$ 27.00 &   3 &  13 & 0.095\\
27.00 $\rightarrow$ 27.50 &   1 &   9 & 0.120\\
\enddata
\label{phottab}
\end{deluxetable}

\begin{deluxetable}{cccccc}
\tablewidth{0in}
\tablecaption{Previously Cataloged Variables}
\tableheadfrac{0.01}
\tablehead{
\colhead{{Name (M31ACSV)}} &
\colhead{{Previous Name}} &
\colhead{{Type}} &
\colhead{{Per. (d)}} &
\colhead{{$B_{DIR}$\tablenotemark{a}}} &
\colhead{{$B_{bright}$}}
}
\tablenotetext{a}{The mean $B$ magnitudes (when available) taken from DIRECT \citep{kaluzny1999}; no $B$ band photometry was available from \citet{joshi2003}.}
\startdata
J004358.1+411250 & [JPN2003] V348 & Irregular & \nodata & \nodata & 23.22$\pm$0.01\\
J004401.0+411132 & DIRECT V4569 M31D & Periodic & 31.4 & 21.66 & 22.13$\pm$0.01\\
J004404.0+411353 & DIRECT V5146 M31D & Cepheid & 13.523 & 22.00 & 22.01$\pm$0.00\\
J004406.6+411122 & [JPN 2003] V357 & Periodic & 132 & \nodata & 24.42$\pm$0.02\\
J004408.0+411243 & DIRECT V5762 M31D & Periodic & 56.6 & \nodata & 22.39$\pm$0.01\\
J004408.4+411347 & DIRECT V5816 M31D & Periodic & 35.1 & 21.64 & 21.92$\pm$0.00\\
\enddata
\label{previous}
\end{deluxetable}

\end{document}